\pdfoutput=1
\documentclass[conference]{IEEEtran}

\usepackage{cite}
\usepackage{amsmath}
\usepackage{amssymb}
\usepackage{amsfonts}
\usepackage{graphicx}
\usepackage{xcolor}
\usepackage{algorithm}
\usepackage{algpseudocode}
\usepackage{textcomp}
\usepackage{booktabs} 
\usepackage{multicol}
\usepackage{multirow}
\usepackage{subfigure}
\usepackage{url}
\usepackage{threeparttable} 
\usepackage{hyperref}
\usepackage{bookmark}

\begin{document}

\title{\Large \bf Unveiling the Bandwidth Nightmare: CDN Compression Format Conversion Attacks}

\author{\IEEEauthorblockN{1\textsuperscript{st} Ziyu Lin}
\IEEEauthorblockA{\textit{Fuzhou University} \\
Fuzhou, China \\
linziyu0205@gmail.com}
~\\
\and
\IEEEauthorblockN{2\textsuperscript{nd} Zhiwei Lin}
\IEEEauthorblockA{\textit{Sichuan University} \\
Chengdu, China \\
 snakinya@gmail.com}

~\\
\and
\IEEEauthorblockN{3\textsuperscript{rd} Ximeng Liu$^\ast$}
\IEEEauthorblockA{\textit{Fuzhou University} \\
Fuzhou, China \\
snbnix@gmail.com}

~\\
\and
\IEEEauthorblockN{4\textsuperscript{th} Zuobing Ying}
\IEEEauthorblockA{\textit{City University of Macau} \\
Macau, China \\
zbying@cityu.mo}

\and
\IEEEauthorblockN{5\textsuperscript{th} Cheng Chen}
\IEEEauthorblockA{\textit{Fuzhou University} \\
Fuzhou, China \\
cc18250259032@163.com}

}

% make the title area
\maketitle

\begin{abstract}
Content Delivery Networks (CDNs) are designed to enhance network performance and protect against web attack traffic for their hosting websites. 
And the HTTP compression request mechanism primarily aims to reduce unnecessary network transfers. 
However, we find that the specification failed to consider the security risks introduced when CDNs meet compression requests. 
In this paper, we present a novel HTTP amplification attack, CDN Compression Format Convert (CDN-Convet) Attacks. 
It allows attackers to massively exhaust not only the outgoing bandwidth of the origin servers deployed behind CDNs but also the bandwidth of CDN surrogate nodes.
We examined the CDN-Convet attacks on 11 popular CDNs to evaluate the feasibility and real-world impacts. 
Our experimental results show that all these CDNs are affected by the CDN-Convet attacks. 
We have also disclosed our findings to affected CDN providers and have received constructive feedback.
\end{abstract}

\section{Introduction}
Content Delivery Networks (CDNs) are regarded as a crucial component of the internet infrastructure, enhancing website performance, scalability, and security by redirecting client users' web requests to geographically distributed proxy servers. 
CDN providers globally deliver web resources, significantly boosting the performance and scalability of hosted websites. 
Furthermore, CDNs are renowned for their intricate protective mechanisms, including filtering or specifying intrusion traffic and offloading distributed denial of service (DDoS) traffic to global proxy nodes. 
As a result, CDN providers enjoy widespread trust among the most popular websites worldwide and find extensive use throughout the internet.

To speed up content delivery, CDNs have developed a compression mechanism. 
This compression mechanism reduces the size of web resources by compressing them at CDN surrogate nodes, resulting in faster transfer speeds. 
Standard compression algorithms, such as Gzip and Brotli, enjoy widespread adoption for the compression of various types of content, including text, images, scripts, and other compressible resources.

In this paper, we present a novel HTTP amplification attack, CDN Compression Format Convert (CDN-Convert) Attacks. 
This attack leverages the asymmetry in bandwidth consumption between client-CDN and CDN-Origin sides caused by the CDN compression mechanism to exhaust the bandwidth resources of the origin server or CDN. 
Attackers can significantly consume the network bandwidth of source servers hosted on the CDN by executing carefully designed HTTP compression requests. 
Simultaneously, using the CDN's caching mechanism, attackers can directly harm the performance of CDN nodes by establishing high-traffic responses between specific CDN nodes without the need for the origin server.

In this study, we also evaluate the CDN-Convert attacks in the wild by conducting a series of controlled experiments on 11 popular CDN vendors. 
Our experiment results show that all examined CDN providers are severely affected by CDN-Convert attacks. 
By crafting HTTP request attacks on the origin server, attackers can force the origin server to generate response traffic that is over 11 times larger than the traffic received by the attacker, achieving amplification effects exceeding 1,000 times when attacking CDN nodes. 
Throughout the research, we placed a strong emphasis on potential ethical concerns. 
Firstly, the victim websites were our own registrations. Secondly, to avoid collateral damage to the CDN platforms, we used only a few dozen CDN nodes to demonstrate the effectiveness of CDN-Convert attacks. 
In reality, we collected millions of available CDN nodes, all of which theoretically could be targeted.
Finally, we discuss the causes of the problem and propose some possible mitigation countermeasures against these attacks. 
We also responsibly disclosed these vulnerabilities to CDN vendors and received positive feedback. Some of which have fixed these vulnerabilities.

\noindent\textbf{Contribution:} In this paper, we make the following contributions:

\begin{itemize}
\item[1] We propose a novel class of HTTP amplification attacks, CDN-Convert attacks, which can consume a victim's outbound bandwidth, reducing network availability and causing economic losses.
\item[2] We examine the CDN-Convert attacks on 11 popular CDN vendors and evaluate the feasibility and severity of compression mechanism vulnerabilities. We find all examined CDNs are vulnerable to CDN-Convert attacks, and the amplification factor is up to 1840 times in some cases.
\item[3] We responsibly disclose all security issues to affected CDN providers. Additionally, we analyze the root causes of CDN-Convert vulnerabilities and propose countermeasures and mitigation solutions.
\end{itemize}

\noindent\textbf{Roadmap:} We organize the rest of this paper as follows. 
In Section~\ref{sec2}, we provide the background of CDN and compression mechanisms. 
In Section~\ref{sec3}, we discuss the compression-specific implementations in CDNs. 
In Section~\ref{sec4}, we describe the details of the CDN-Convert attacks. 
We also evaluate the feasibility of the CDN-Convert attacks and explore the amplification factors in Section~\ref{sec5}. 
We discuss mitigation solutions and our responsible disclosure in Section~\ref{sec6}. 
Section~\ref{sec7} elaborates on the related works, including CDN security and amplification attacks. 
We conclude in Section~\ref{sec8}.
\section{Background}\label{sec2}
In this section, we provide a brief introduction to the essential features of CDN and the compression mechanisms employed within CDN.

\subsection{CDN Overview}\label{subsec2.1}
CDN is a network infrastructure, possesses multiple important functions aimed at improving website performance and protecting data security. 
A key role of CDN lies in its ability to conserve bandwidth at the origin servers. By caching both static and dynamic content of websites on edge nodes, CDN can directly serve content from nearby edge nodes when users request access to the website. 
This reduces the need for content transmission from the origin servers, eliminating long-distance transfers and effectively saving bandwidth and traffic. 
Moreover, CDN leverages compression algorithms to compress requested content, reducing file sizes and further minimizing data transfer volume and bandwidth consumption. 
Through these measures, CDN significantly reduces website traffic costs while providing a more cost-effective and efficient content delivery service.

\begin{figure}
    \centering
    \caption{Multiple segments of connectivity in a CDN environment}
    \includegraphics[width=1.0\linewidth]{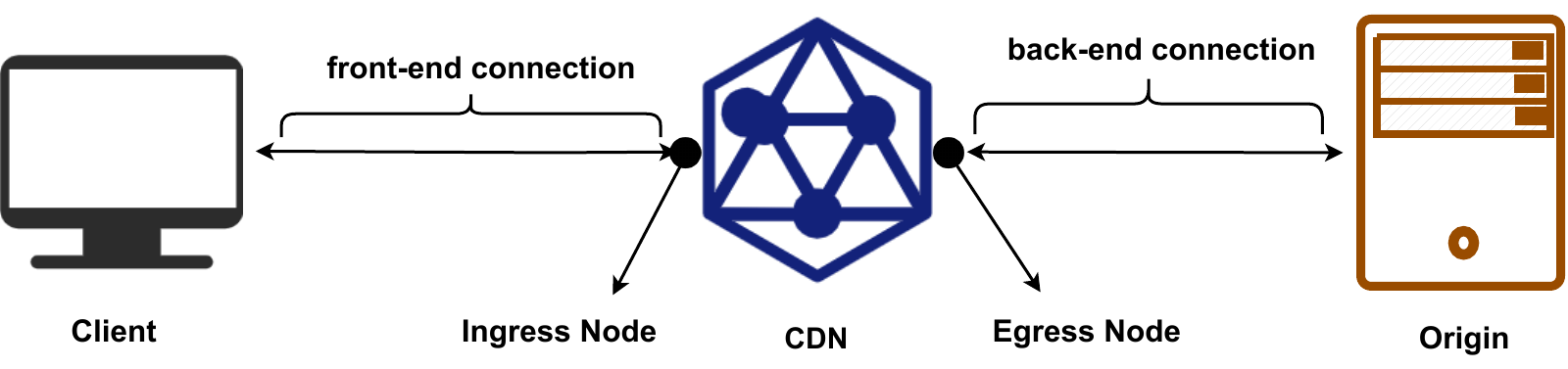}
    \label{fig1:env}
\end{figure}

As shown in Fig. \ref{fig1:env}, a typical CDN network comprises central nodes and edge nodes~\cite{HeterChannels}. 
The central nodes are responsible for global load balancing and content management, while the edge nodes are dedicated to content distribution and caching. 
The edge nodes are further categorized into entry nodes and exit nodes based on their location and functionality. 
Generally, entry nodes are situated in close proximity to users, handling user access requests and content distribution. 
On the other hand, exit nodes are located near the source website and are responsible for retrieving content from the source website.

By employing cascading techniques, we can achieve the concatenation of CDNs~\cite{Marco}, as depicted in Fig.~\ref{fig2:threat_model}. 
In this configuration, we refer to the CDN closer to the user side as UCDN and the CDN closer to the origin server side as OCDN. 
When a user requests data, the CDN first attempts to respond from its local cache~\cite{CDN_useage}. 
If the local cache misses, the CDN forwards the request to the origin server to retrieve the target resource and caches the response for subsequent requests. 
This mechanism effectively reduces user access latency and alleviates the load on the origin server. 
Additionally, CDN provides origin server DDoS protection by dynamically selecting edge nodes through load balancing~\cite{CDN_analysis}.

However, users can intentionally trigger cache misses and force web requests to be forwarded to the origin server. 
Previous studies~\cite{decoupleAmp, CDN-Convex,bypassCache} has shown that CDN vendors simply forward any requests with a WebSocket handshake header to the origin server, and CDN's caching mechanism can also be exploited by appending a random query string to the target URL. 
Furthermore, many CDNs offer configurable options to customize caching policies, enabling malicious clients to disable resource caching.

\subsection{Compression Mechanism on CDN}\label{subsec2.2}
CDN compression technology~\cite{compress} is widely employed to optimize the delivery of web page content to end users. 
This technology leverages compression algorithms to reduce the size of web resources, thereby minimizing data transmission and bandwidth consumption. 
Given the escalating demand for high-quality content and faster loading times, CDN compression plays a pivotal role in facilitating the efficient operation of the internet.

Two commonly utilized compression algorithms are Gzip and Brotli. Gzip~\cite{gzip}, developed by the GNU project, enjoys extensive support and usage for compressing web page content. 
It employs the Deflate compression algorithm, which combines the LZ77 algorithm and Huffman coding, rendering it particularly effective for compressing textual content such as HTML, CSS, and JavaScript files. 
Conversely, Brotli~\cite{Brotli}, a more recent compression algorithm developed by Google, offers superior compression ratios compared to Gzip. 
This implies that Brotli can achieve smaller file sizes, resulting in faster transmission times. 
Brotli exhibits exceptional efficacy when compressing textual and mixed content, and it is supported by the majority of modern web browsers.

These compression algorithms can be applied to any compressible resource, encompassing text, images, scripts, and more. 
The selection of the algorithm and compression level can be tailored to the specific requirements of web content and the capabilities of client devices. 
During the transmission process, the CDN automatically detects whether the client supports compression. 
If compression is supported, the CDN compresses the content prior to transmission. 
This functionality is facilitated through the Accept-Encoding HTTP header, which informs the server about the compression algorithms supported by the client.

The compression process not only conserves bandwidth and reduces data traffic but also significantly enhances website loading speed, thereby augmenting the overall user experience. 
By reducing file sizes, CDN compression enables expedited download speeds and mitigates user waiting times, particularly for users with slower network connections. 
This, in turn, enhances user engagement, diminishes bounce rates, and optimizes website performance.
\section{Compression-Specific Implementations In CDNs}\label{sec3}
In this section, we first present why we specifically explore these 11 CDN vendors. Then, we analyze and clarify their compression request handling behaviors which lead to the CDN-Convert attacks.

\subsection{Consideration in Selecting CDN Vendors}\label{subsec3.1}
We test 11 popular CDNs around the world, including Azure, Alibaba Cloud, Bunny, Baidu Cloud, CDN77, Cloudflare, CloudFront, CDNetworks, G-core, and Tencent. 
These CDNs are often studied in previous related works~\cite{CDN-Abuse,cdn-loop} and most of them rank high in the market share~\cite{CDN-market}. 
Moreover, most of these CDNs provide free or free-trial accounts, which indicates little cost to launch an attack. 
Baidu does not offer a free service, and we have spent less than $\$1$ in our experiment. 
Tencent and Alibaba only offer paid services, but give 50GB of free traffic per month for six months. 
In all subsequent experiments, we deploy our origin server individually behind these CDNs.

\subsection{Differences in CDNs Handling Compression Requests}\label{subsec3.2}

To determine which CDNs support compression requests, we configure the origin servers that do not support compression and then send various valid compression requests to each CDN. 
Experimental results show that our origin servers return responses without the Vary: Accept-Encoding header, but all CDNs always return responses with the Vary: Accept-Encoding header for gzip or brotli compression requests. 
Therefore, we conclude that all 11 CDNs support compression of responses on edge servers.

Different CDN vendors choose to implement Accept-Encoding forward policies based on different perspectives, including business, operational, and technical views. 
At present, there are three basic Accept-Encoding requests forward policies deployed on CDN, including:

\begin{itemize}
\item[1] Laziness - Forward the Accept-Encoding header without change. 
\item[2] Deletion - Remove the Accept-Encoding header directly. 
\item[3] Modification - Modify the Accept-Encoding header.
\end{itemize}

As shown in Table~\ref{tab1}, most CDNs tend to adopt either the deletion policy or the modification policy when forwarding compression requests to the origin server for various reasons, such as mitigating potential zip bomb attacks, minimizing origin server load, and improving performance.

\begin{table*}[t]
    \caption{CDNs' Accept-Encoding Modification Behavior}
    \centering
    \small
    \begin{threeparttable}
    \begin{tabular}{cccccc}
            \toprule[1.5pt]
                   & \multicolumn{5}{c}{Accept-Encoding}                                                               \\
                   & gzip              & compress          & deflate           & br                & identity          \\
            \midrule
        Azure      & gzip              & compress          & defalte           & br                & delete            \\
        Alibaba    & gzip              & compress          & defalte           & br                & identity          \\
        Bunny      & gzip, deflate, br & gzip, deflate, br & gzip, deflate, br & gzip, deflate, br & gzip, deflate, br \\
        Baidu      & gzip              & delete            & defalte           & br                & delete            \\
        CloudFront & gzip              & compress          & defalte           & br                & identity          \\
        Cloudflare & gzip              & gzip              & gzip              & gzip              & gzip              \\
        CDN77      & delete            & delete            & delete            & delete            & delete            \\
        CDNetworks & gzip              & delete            & delete            & delete            & delete            \\
        G-core      & delete            & delete            & delete            & delete            & delete            \\
        Tencent    & gzip              & delete            & defalte           & br                & identity          \\
        UPYun      & delete            & delete            & delete            & delete            & delete            \\
            \bottomrule[1.5pt]
            \end{tabular}
    \end{threeparttable}
    \label{tab1}
\end{table*}
\section{CDN Convert Format Attack}\label{sec4}
The Deletion and Modification policies are beneficial for CDNs to improve service performance. 
However, we note that these policies require CDNs to retrieve uncompressed data from the origin server, which causes the origin server to consume more traffic. 
Also, if the CDN decompresses the compressed response, the response sent by the CDN can be thousands of times larger than the one from the origin server. 
These cases will cause severe traffic differences between different connections in the network path from the client to the origin server.

\subsection{Threat Model}\label{subsec4.1}

In this study, we assume that an adversary has two limited capabilities of attackers. 
Firstly, we assume an attacker is able to craft malicious but legal requests like a benign end-user to the CDN. 
Secondly, the attacker can register an account with CDN providers and host the victim's website on the CDN. 
Currently, many CDN providers, likely driven by competitive factors, offer free or free-trial services to potential customers (including attackers) without strong identity verification~\cite{cdn-loop}. 
Consequently, attackers can initiate attacks anonymously at a low cost. 

From the perspective of potential victims, any website could potentially be affected. 
In today's CDN operations, there is often a leniency in customer-controlled forwarding policies, and a lack of robust source verification~\cite{CDN-Abuse}. 
This implies that a malicious CDN customer has the capability to set up CDN edge servers to redirect traffic to any domain name or IP address of their choosing, regardless of ownership.

\begin{figure}
    \centering
	\caption{General construction of the CDN-Convert Attacks}
    \includegraphics[width=1.0\linewidth]{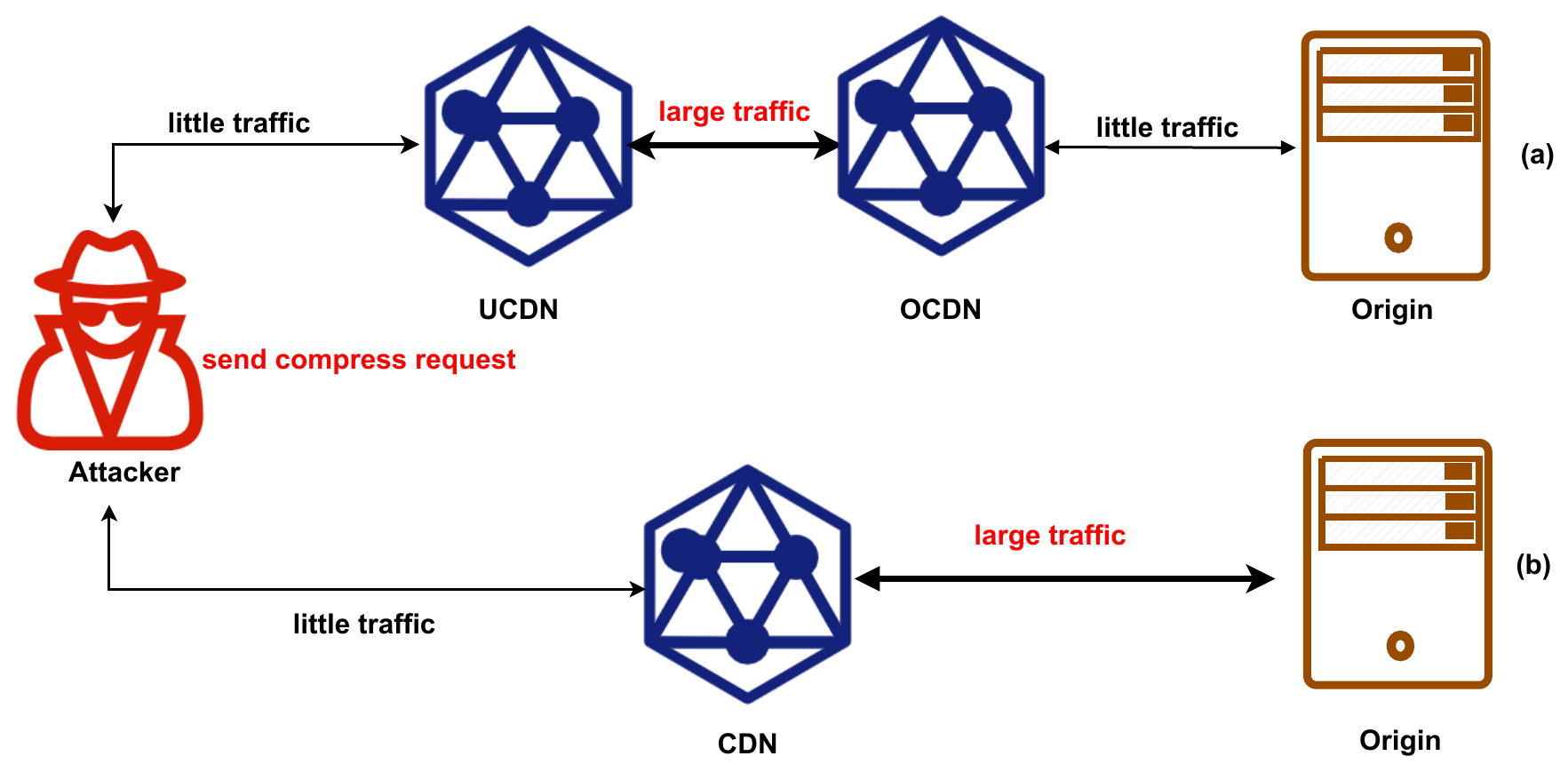}
    \label{fig2:threat_model}
\end{figure}

The significant traffic differences caused by file format conversion will bring a novel class of traffic amplification attacks, which we refer to as "CDN Compression Format Conversion (CDN-Convert) attacks." 
We identify two scenarios of CDN-Convert attacks, discussed separately in Section~\ref{subsec4.2} and Section \ref{subsec4.3}.
In a CDN-Convert attack, the attacker is able to craft malicious but legal requests to the CDN, as shown in Fig.~\ref{fig2:threat_model} One of the victims is the origin server in Fig. 2a, which is being normally hosted on the CDN by the owner, or maliciously deployed on the CDN by the attacker~\cite{CDNJudo}. 
The other victims are the UCDN and the OCDN in Fig. 2b, which are maliciously cascaded together by the attacker. 

Through an empirical study, we show that the attacker can perform a traffic amplification attack with little cost and exhaust the bandwidth of its victims.

\subsection{CDN Convert to Compression Format (CCCF) Attack}\label{subsec4.2}

If the CDN use Modification or Deletion policy to handle requests, an attacker can perform a CCCF attack by including the accept-encoding header in the request to the origin server. 
Through this attack, the CDN-origin connection will transfer much larger traffic than the client-CDN connection, allowing the attacker to target the origin server hosted on the CDN.

\begin{figure}[h]
    \centering
	\caption{Flow and example construction of a CCCF attack.}
    \includegraphics[width=1\linewidth]{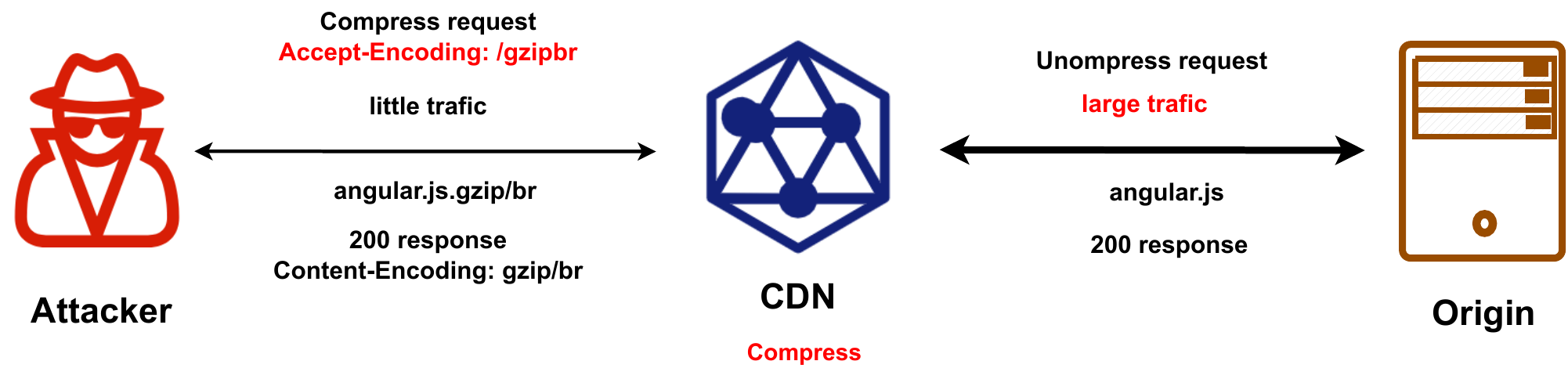}
    \label{fig3:CCCF_attack}
\end{figure}

As shown in Fig.~\ref{fig3:CCCF_attack}, the attacker can include the request header accept-encoding: gzip or accept-encoding: br when making a request to a vulnerable CDN. 
The CDN will process this request header and then forward it to the origin server. The processing may involve converting accept-encoding: br to accept-encoding: gzip or removing the request header altogether. 
This results in the origin server returning an uncompressed file or a file with a smaller compression ratio, while the CDN returns a file with a higher compression ratio.
Even worse, some CDNs can modify the request header. 
For CDNs that do not adopt the deletion policy, we can use this feature to delete the accept-encoding request header and force the origin server to return uncompressed resources.

\subsection{CDN Convert to Uncompressed Format (CCUF) Attack}\label{subsec4.3}
If the UCDN and OCDN adopt different policies to handle requests and the OCDN supports decompression, then an attacker can carry the Accept-Encoding request header to launch a CCUF attack. 
This attack comes in two types, in which the UCDN-OCDN connection will transmit much larger traffic than the OCDN-origin connection, allowing the attacker to consume a large amount of available bandwidth between UCDN and OCDN. 
The attacker can send multiple requests to the same ingress node of UCDN and cascade UCDN and OCDN to launch a CCUF attack on a specific ingress node of UCDN.

\begin{figure*}[ht]
    \centering
    \subfigure[CCUF Type1]{\includegraphics[width=0.48\textwidth]{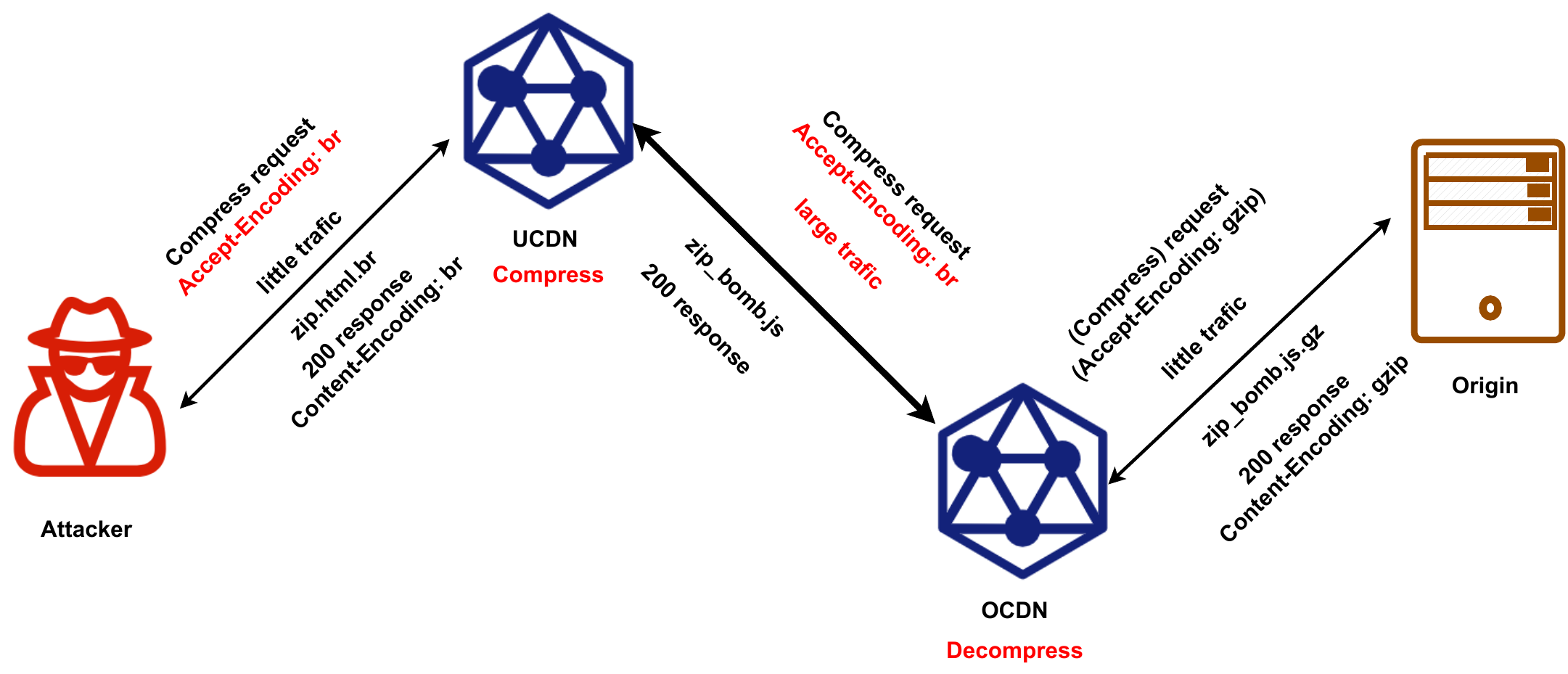}}\hspace{0.5cm}
    \subfigure[CCUF Type2]{\includegraphics[width=0.48\textwidth]{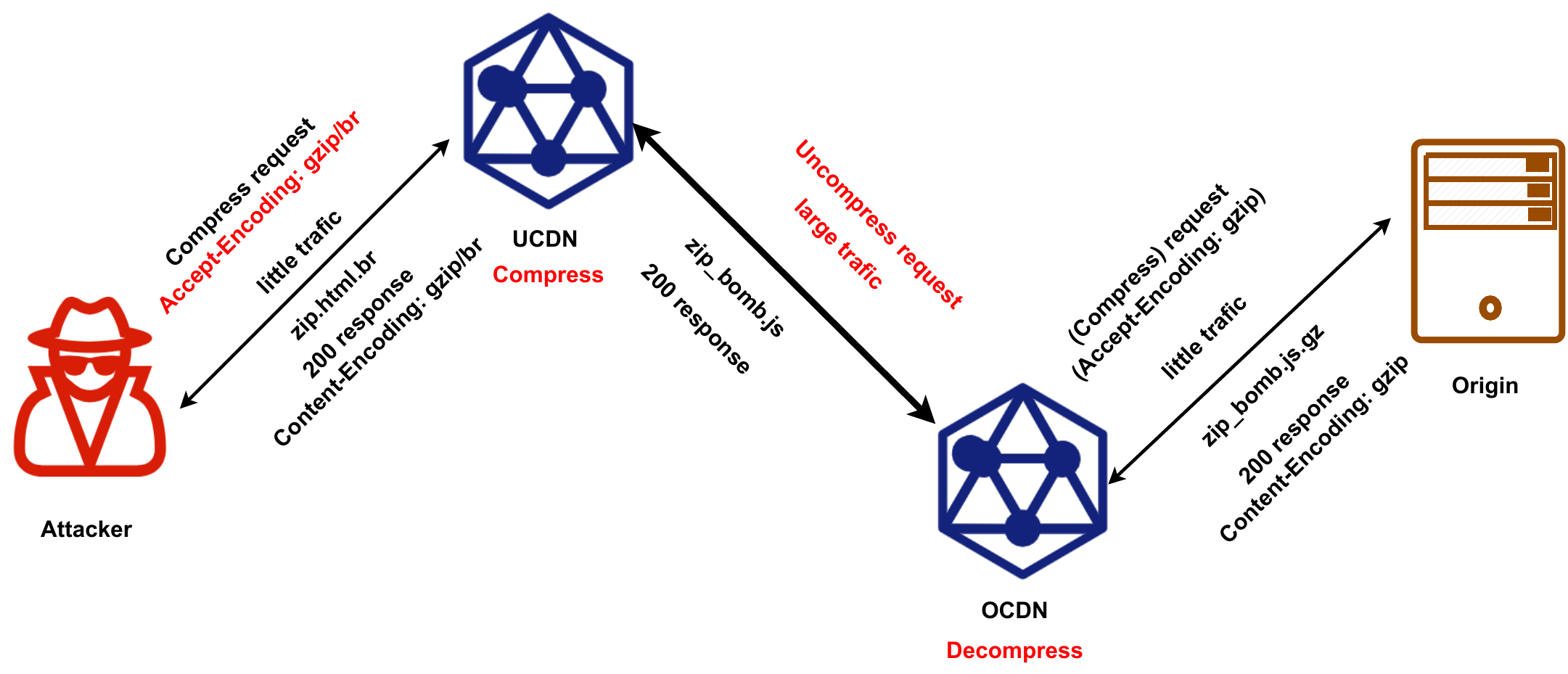}}
    \caption{Two new ways of domain takeover.}
    \label{fig4:CCUF_attack} 
\end{figure*}

In the first type of CCUF attack, as shown in Fig.~\ref{fig4:CCUF_attack}, the attacker carries the request header Accept-Encoding: br to request a file from a vulnerable CDN. 
When handling the compression request, UCDN adopts the Laziness policy, which means it does not modify the request and forwards it to OCDN. Upon receiving the request, OCDN adopts the Deletion policy or the Modification policy, changing Accept-Encoding: br to Accept-Encoding: gzip. 
The origin server responds to OCDN with a gzip-compressed file.
Since OCDN supports decompression and does not support Brotli compression under HTTP, it will decompress the gzip'd response and pass it to UCDN. 
UCDN then compresses the response using Brotli. 
This results in the traffic between the attacker-UCDN and OCDN-origin being much smaller than that between UCDN-OCDN.

In the second type of attack, as shown in Fig.~\ref{fig4:CCUF_attack}, the attacker carries the request header Accept-Encoding: gzip or br to request a gzip bomb from the vulnerable CDN. 
UCDN adopts the Deletion strategy in processing the request, which means it deletes the request header before forwarding it to OCDN. 
OCDN will decompress the corresponding response for requests without the Accept-Encoding header. Since the origin server is under the attacker's control, we let it respond to OCDN with a gzip bomb. 
Then OCDN will decompress the gzip bomb and pass it to UCDN. 
UCDN then compresses the bomb using gzip or brotli. 
This results in the traffic between the attacker-UCDN and OCDN-origin being much smaller than that between UCDN-OCDN.

\section{Real-world Evaluation}\label{sec5}
To explore the feasibility and severity of CDN-Convert vulnerabilities in the wild, we conducted a series of experiments. 
We examine which CDNs are vulnerable to CDN-Convert attacks, calculate the actual amplification factors, and analyze the practical impacts. 
In all experiments, our origin server is the same Linux server with 2.4GHz of CPU, 4G of memory, and 30Mbps of bandwidth. 

\subsection{Feasibility of the CCCF Attacks}\label{subsec5.1}
In the first experiment, our origin server is a default configuration of Nginx/1.21.3. 
We sent a series of compression requests to each CDN, ensuring that they were forwarded to our origin server. 
Simultaneously, we collected all requests and responses on both the client and origin server sides.
We compared the requests sent by the clients with the corresponding requests received by the origin server to analyze how each CDN modified the Accept-Encoding header. 
Additionally, we compared the responses sent by the origin server with those received by the clients to assess the CDN's support for various compression algorithms. 
Our experiment results show that all 11 major CDNs were vulnerable to CCCF attacks. 
As shown in Table~\ref{tab1}, how CDNs handled the Accept-Encoding header under default configuration. 
As shown in Table~\ref{tab2}, all 11 CDNs supported the compression of responses on their edge servers. 
The second column lists the supported compression modes, the third column specifies the supported compression algorithms, and the fourth column indicates the compression levels. 
The details are as follows:

\begin{table}[ht]
    \centering
    \caption{CDNs' Compression Behavior}
    \small
    \begin{threeparttable}
    \begin{tabular}{cccc}
        \toprule[1.5pt]
        CDN & Compression Mode & Algorithms & Level \\
        \midrule
        Azure & Edge server and origin & gzip br & gzip5 br5 \\
        Alibaba & Edge server and origin & gzip br & gzip5 br1 \\
        Bunny & Edge server and origin & gzip br & gzip2 br2 \\
        Baidu & Edge server and origin & gzip br & gzip4 br4 \\
        CloudFront & Edge server and origin & gzip br & gzip2 br6 \\
        Cloudflare & Edge server and origin & gzip br & gzip2  br4 \\
        CDN77 & Edge server & gzip br & gzip5 \\
        CDNetworks & Edge server and origin & gzip & gzip3 \\
        Gcore & Edge server & gzip & gzip2 \\
        Tencent & Edge server and origin & gzip br & gzip5 br5 \\
        UPYun & Edge server & gzip br & gzip5 br5 \\
        \bottomrule[1.5pt]
    \end{tabular}
    \end{threeparttable}
    \label{tab2}
\end{table}

Azure, Alibaba Cloud, Baidu Cloud, CloudFront, and Tencent adopted the Lazines policy for handling the Accept-Encoding header. 
But if the origin server does not support compression, they will compress the response on the edge server. 
Therefore, attackers can configure CDNS to delete the Accept-Encoding header to abuse them to launch an CCCF attack.

Cloudflare adopted a modification policy for handling the Accept-Encoding header. 
Specifically, irrespective of whether the origin server supported compression, Cloudflare set all Accept-Encoding headers to gzip and requested content in gzip-compressed format from the origin server. 
Subsequently, the response content is converted at the edge server into the client's desired format. 
Given Cloudflare's ability to modify HTTP request headers, attackers could configure Cloudflare to remove the Accept-Encoding header, compelling the origin server to provide uncompressed resources. 
The CDN would then compress these resources and deliver them to the attacker, which can be abused to launch a CCCF attack.

CDN77, G-Core, and UpYun adopted the deletion policy for Accept-Encoding headers. 
While this prevents potential zip bomb attacks, it also results in much greater traffic consumption on the CDN-origin side than on the client-CDN side. 
An attacker could abuse this to launch a CCCF attack.

Cdnetworks adopted the laziness and modification policy for handling the Accept-Encoding header.
It only forwards Accept-Encoding: gzip requests and supports the removal of HTTP headers. 
Attackers can configure Cdnetworks to remove Accept-Encoding so that the origin server will only return uncompressed resources, which they can exploit to launch a CCCF attack.

\subsection{The Amplification Factor of the CCCF Attack}\label{subsec5.2}
\textbf{Experiments Setup.} We registered the service of 11 CDN vendors and set up an HTTP service as the victim on a cloud VPS (2.5GHz/4GB/30Mbps) located in Singapore. 

\textbf{Experiments Result.} We found that the amplification factor of the CCCF attack is related to the compression algorithm, so we sent GET requests with Accept-Encoding headers specifying either gzip or brotli to each CDN. As shown in Table~\ref{tab3}, our experiment results show that all 11 prominent CDNs are vulnerable to CCCF attacks, and CDNs that support brotli compression are more vulnerable to CCCF attacks.

\begin{table*}[ht]
    \centering
    \caption{The Amplification Factor Of The CCCF Attack With angular.js}
    \small
    \begin{threeparttable}
        \begin{tabular}{ccccc}
            \toprule[1.5pt]
            CDN Vendor & Exploited Case    & Client-CDN Traffic Size & CDN-Origin Traffic Size & Amplification Factor \\
            \midrule
            Azure      & Accept-Encoding: gzip & 53896B                   & 333247B                  & 6.18          \\
            Alibaba    & Accept-Encoding: gzip & 55095B                   & 332973B                  & 6.04          \\
            Bunny      & Accept-Encoding: gzip & 62981B                   & 333037B                  & 5.29          \\
            Baidu      & Accept-Encoding: gzip & 35840B                   & 333042B                  & 9.29         \\
            CloudFront & Accept-Encoding: gzip & 53818B                   & 332845B                  & 6.18         \\
            Cloudflare & Accept-Encoding: gzip & 53627B                   & 332850B                  & 6.21          \\
            CDN77      & Accept-Encoding: gzip & 53213B                   & 339968B                  & 6.39         \\
            CDNetworks & Accept-Encoding: gzip & 61018B                   & 332700B                  & 5.45          \\
            Gcore      & Accept-Encoding: gzip & 63963B                   & 332946B                  & 5.21          \\
            Tencent    & Accept-Encoding: gzip & 53666B                   & 332897B                  & 6.20          \\
            UPYun      & Accept-Encoding: gzip & 54457B                   & 332846B                  & 6.11         \\
            Azure         & Accept-Encoding: br & 44388B                   & 333245B                  & 7.51          \\
            Alibaba           & Accept-Encoding: br & 86553B                   & 332970B                  & 3.85          \\
            Bunny             & Accept-Encoding: br & 53381B                   & 333037B                  & 6.24          \\
            Baidu             & Accept-Encoding: br & 30720B                   & 333042B                  & 10.84                 \\
            CloudFront        & Accept-Encoding: br & 44667B                   & 332843B                  & 7.45          \\
            Cloudflare        & Accept-Encoding: br & 47104B                   & 332850B                  & 7.07          \\
            Tencent           & Accept-Encoding: br & 47555B                   & 332894B                  & 7.00          \\
            UPYun             & Accept-Encoding: br & 45323B                   & 332846B                  & 7.34          \\
            \bottomrule[1.5pt]
            \end{tabular}  
    \end{threeparttable}
    \label{tab3}
\end{table*}

Even worse, we find instances of websites that permit user-uploaded files, a potential avenue for attackers to exploit. 
In particular, an attacker could upload a zip bomb and leverage CCCF to request this malicious payload, leading to a substantially magnified impact.
For ethical reasons, we only test the amplification factor of a 1MB zip bomb, as shown in Table~\ref{tab4}.

\begin{table*}[ht]
    \centering
    \caption{The Amplification Factor Of The CCCF Attack With 1MB gzip Bomb}
    \small
    \begin{threeparttable}
        \begin{tabular}{ccccc}
            \toprule[1.5pt]
            CDN Vendor & Exploited Case    & Client-CDN Traffic Size & CDN-Origin Traffic Size & Amplification Factor \\
            \midrule
            Azure      & Accept-Encoding: gzip & 5265B                    & 1049638B                 & 199.36          \\
            Alibaba    & Accept-Encoding: gzip & 3845B                    & 1049335B                 & 272.91           \\
            Bunny      & Accept-Encoding: gzip & 7355B                    & 1049418B                 & 142.68          \\
            Baidu      & Accept-Encoding: gzip & 3212B                    & 1049425B                 & 326.72          \\
            CloudFront & Accept-Encoding: gzip & 2270B                    & 1049226B                 & 462.21          \\
            Cloudflare & Accept-Encoding: gzip & 2009B                    & 1049237B                 & 522.27          \\
            CDN77      & Accept-Encoding: gzip & 1503B                    & 1048890B                  & 697.86                     \\
            CDNetworks & Accept-Encoding: gzip & 5301B                    & 1049097B                 & 197.91          \\
            Gcore      & Accept-Encoding: gzip & 5157B                    & 1049327B                 & 203.48         \\
            Tencent    & Accept-Encoding: gzip & 1630B                    & 1049279B                 & 643.73          \\
            UPYun      & Accept-Encoding: gzip & 1784B                    & 1049227B                 & 588.13          \\
            Azure      & Accept-Encoding: br & 825B                     & 1049628B                 & 1272.27          \\
            Alibaba    & Accept-Encoding: br & 2922B                    & 1049352B                 & 359.12           \\
            Bunny      & Accept-Encoding: br & 2944B                    & 1049418B                 & 356.45          \\
            Baidu      & Accept-Encoding: br & 2955B                    & 1049425B                 & 355.13          \\
            CloudFront & Accept-Encoding: br & 818B                     & 1049224B                 & 1282.66          \\
            Cloudflare & Accept-Encoding: br & 967B                     & 1049231B                 & 1085.03          \\
            Tencent    & Accept-Encoding: br & 570B                     & 1049276B                 & 1840.83         \\
            UPYun      & Accept-Encoding: br & 994B                     & 1049227B                 & 1055.56         \\
            \bottomrule[1.5pt]
            \end{tabular}  

    \end{threeparttable}
    \label{tab4}
\end{table*}

\subsection{Feasibility of the CCUF Attacks}\label{subsec5.3}
In the second experiment, we configured an origin server powered by nginx/1.21.3 in two distinct ways. 
One configuration, defined as gzip static always, forced responses to be compressed with gzip, while the other, denoted as brotli static always, mandated responses to be compressed with brotli. 
We sent Accept-Encoding identity requests to each CDN, ensuring they were forwarding to our origin server. 
Simultaneously, we used the tcpdump tool to capture all requests and responses on both the client and origin server sides. 
We compared the responses sent by the origin server with those received by the clients to analyze the CDNs' decompression behavior. 
Our findings indicate that four CDNs (Bunny, Cloudflare, Cdnetworks, Upyun) decompressed gzip responses for clients that did not support compression encoding. 
Nevertheless, none of these CDNs supported decompression for brotli compression.

\subsection{The Amplification Factor of the CCUF Attack}\label{subsec5.4}
To verify the feasibility and effectiveness of the CCUF attacks, we conducted two experiments. 
To minimize or avoid causing actual performance impact on the CDN vulnerable to attacks, we opted to use a 1MB zip bomb containing only the character ``0'' and compressed it using gzip at compression level 9, resulting in a compression ratio of 1,064. 
Additionally, we configured the origin server to only return the gzip'd bomb. 
Furthermore, we set up a proxy server between UCDN and OCDN to collect the traffic transmitted between them. 
To achieve this, we configured UCDN's source server as our proxy server and set up the proxy server to forward requests to OCDN. 
Ultimately, we captured all the traffic transmitted through the client-UCDN connection and the UCDN-OCDN connection and calculated the amplification factor.

\textbf{The Setup of the First Experiment.} Based on the preceding analysis, it is evident that 6 CDNs can be exploited as UCDN, and 2 CDNs can be exploited as OCDN. 
Consequently, the first type of CCUF attack could potentially impact 12 distinct CDN cascading combinations.

\textbf{Results of the First Experiment.} As shown in Table~\ref{tab5}, the traffic for the UCDN-client connection remains under 5000 bytes, but the response traffic for the OCDN-UCDN connection is significantly larger. 
For instance, when abusing Baidu as the UCDN and Cloudflare as the OCDN, the maximum amplification factor is approximately 259. 
Similarly, when utilizing Tencent as the UCDN and Cdnetworks as the OCDN, the maximum amplification factor is approximately 1475. 
However, due to Cloudflare's inability to remove its CDN identity from its response headers, four UCDNs are unable to compress their responses. 
Consequently, only eight distinct CDN cascading combinations are vulnerable to the first type of CCUF attack.

\begin{table*}[ht]
    \centering
    \caption{The Amplification Factor Of The CCUF Attack Type1}
    \small
    \begin{threeparttable}
        \begin{tabular}{cccccc}
            \toprule[1.5pt]
            &            &              & \multicolumn{3}{c}{Exploiting with 1MB of Zip Bomb}                                   \\
            UCDN       & OCDN & Exploited Case & Client-UCDN Traffic Size & UCDN-OCDN Traffic Size & Amplification Factor \\
            \midrule
            Alibaba    & Cloudflare & Accept-Encoding: br           & 4911B                       & 1049802B                  & 213.77               \\
            Baidu      & Cloudflare & Accept-Encoding: br           & 4048B                       & 1049794B                  & 259.34               \\
            Azure      & Cdnetworks & Accept-Encoding: br           & 996B                        & 1049951B                  & 1054.17              \\
            Alibaba    & Cdnetworks & Accept-Encoding: br           & 4068B                       & 1049423B                  & 257.97               \\
            Bunny      & Cdnetworks & Accept-Encoding: br           & 2829B                        & 1049817B                  & 371.10              \\
            Baidu      & Cdnetworks & Accept-Encoding: br           & 816B                        & 1049508B                  & 1286.16              \\
            Cloudfront & Cdnetworks & Accept-Encoding: br           & 1452B                       & 1049381B                  & 722.71               \\
            Tencent    & Cdnetworks & Accept-Encoding: br           & 711B                        & 1049404B                  & 1475.95              \\
            \bottomrule[1.5pt]
            \end{tabular}  
    \end{threeparttable}
    \label{tab5}
\end{table*}

\textbf{The Setup of the Second Experiment.} According to the preceding analysis, 3 CDNs can be abused as the UCDN and 4 CDNs can be abused as the OCDN. 
As a result, the second type of CCUF attack may affect 12 different CDN cascading combinations.

\textbf{Results of the Second Experiment.} As illustrated in Table~\ref{tab6}, the traffic for the UCDN-client connection remains under 2000 bytes, except for G-core. 
However, the traffic for the OCDN-UCDN connection consistently exceeds 1MB. 
For instance, when exploiting CDN77 as the UCDN and Cdnetworks as the OCDN, the maximum amplification factor is approximately 590. 
Similarly, when abusing Upyun as the UCDN and Cloudflare as the OCDN, the attacker requests a br-compressed zip bomb from Upyun, resulting in a maximum amplification factor of approximately 969.

\begin{table*}[ht]
    \centering
    \caption{The Amplification Factor Of The CCUF Attack Type2}
    \small
    \begin{threeparttable}
        \begin{tabular}{cccccc}
            \toprule[1.5pt]
            &            &              & \multicolumn{3}{c}{Exploiting with 1MB of Zip Bomb}                            \\
        UCDN   & OCDN       & Exploited Case & Client-UCDN Traffic Size & UCDN-OCDN Traffic Size & Amplification Factor \\
            \midrule
        \multirow{4}{*}{CDN77}  & Bunny      & Accept-Encoding: gzip         & 1894B                        & 1049791B                   & 554.27                  \\
               & Cloudflare & Accept-Encoding: gzip         & 1835B                        & 1049658B                   & 572.02                  \\
               & Cdnetworks & Accept-Encoding: gzip         & 1820B                        & 1049469B                   & 576.63                  \\
               & Upyun      & Accept-Encoding: gzip         & 1859B                        & 1049529B                   & 564.57                  \\
        
        \multirow{4}{*}{G-core} & Bunny      & Accept-Encoding: gzip         & 5708B                        & 1049885B                   & 183.93                  \\
               & Cloudflare & Accept-Encoding: gzip         & 5312B                        & 1049760B                   & 197.62                  \\
               & Cdnetworks & Accept-Encoding: gzip         & 5299B                        & 1049474B                   & 198.05                  \\
               & Upyun      & Accept-Encoding: gzip         & 5406B                        & 1049581B                   & 194.15                  \\
        
        \multirow{4}{*}{Upyun}  & Bunny      & Accept-Encoding: gzip         & 2242B                        & 1049789B                   & 468.24                  \\
               & Cloudflare & Accept-Encoding: gzip         & 1899B                        & 1049667B                   & 552.75                  \\
               & Cdnetworks & Accept-Encoding: gzip         & 1919B                        & 1049337B                   & 546.84                  \\
               & Upyun      & Accept-Encoding: gzip         & 1881B                        & 1049466B                   & 557.93                  \\
        
               & Bunny      & Accept-Encoding: br           & 1209B                        & 1049789B                   & 868.31                  \\
               & Cloudflare & Accept-Encoding: br           & 1109B                        & 1049667B                   & 946.50                  \\
               & Cdnetworks & Accept-Encoding: br           & 1129B                        & 1049337B                   & 929.47                  \\
               & Upyun      & Accept-Encoding: br           & 1091B                        & 1049483B                   & 961.95                  \\
               \bottomrule[1.5pt]
            \end{tabular}  
    \end{threeparttable}
    \label{tab6}
\end{table*}

\subsection{Severity Assessment}\label{subsec5.5}

\textbf{A serious and common practical impact.} According to our experimental results, the amplification factor of the CCCF attack is almost proportional to the compression level, while the amplification factor of the CCUF attack is proportional to the gzip bomb compression rate.  
All 11 CDNs we tested are vulnerable to the CCCF attack, and 20 combinations of cascaded CDNs are vulnerable to the CCUF attack. As we described in Section~\ref{subsec3.1}, these CDNs are popular around the world and rank high in the market share. 
Thus, there are lots of websites and CDN nodes exposed to our CDN-Convert vulnerability.

\textbf{A low-cost and efficient DDoS attack.} Unlike other DDoS attacks that need to control a large scale of botnets~\cite{Botnet-DDoS}, the attacker only needs an ordinary laptop to launch the CDN-Convert attacks. 
The ingress nodes of CDNs are scattered around the world, coming into a natural distributed ``botnet''. 
This makes a CDN-Convert attacker able to easily congest the target network and even create a denial of service in seconds, while the attacker pays a small cost. 

\textbf{A great monetary loss to the victims.} Most CDNs charge their website customers by traffic consumption, including Alibaba Cloud, Azure, Bunny, Baidu Cloud, CDN77, Cloudflare, CloudFront, G-core, Tencent Cloud~\cite{Alibaba-price, Azure-price, Tencent-price, CloudFront-price}.
When a website is hosted on a vulnerable CDN, its opponent can abuse the CDN to perform a CDN-Convert attack against it, causing a very high CDN service fee to the website. 

\textbf{A security challenge to anti-DDoS.} Traditional DDoS attacks that consume bandwidth mainly target the victim's incoming bandwidth. 
Instead, The CDN-Convert attacks mainly consume the victim’s outgoing bandwidth. This will pose a security challenge to the detection of DDoS attacks. 
\section{Discussion}\label{sec6}
\subsection{Ethic Consideration}\label{subsec6.1}

Our goal has been to keep a balance between real-world severity evaluations and the risks of affecting CDN providers throughout our research. 
Our main worry was the likelihood that higher bandwidth usage during our tests would impair the CDN's network performance and harm other websites hosted on the CDN. 
Therefore, we have taken great care to ensure that no ethical problems arise during our experiments. 
Firstly, we conduct controlled experiments to limit bandwidth consumption in both time and volume dimensions. 
Secondly, in the experiments for the CCCF attack, we send only one compression request to the CDN at a time, which hardly affects the performance of the CDN, and the origin website is implemented by ourselves. 
Thirdly, in the experiments for the CCUF attack, our gzip bomb decompressed to a size of only 1MB does not generate excessive traffic in the Ucdn-Ocdn connection and we send all requests to completely different entry nodes of the CDN to minimize or avoid any real impact on the performance of a specific node. 
Hence, we firmly believe that the benefits of our work outweigh any harm caused.

\subsection{Mitigations}\label{subsec6.2}
\textbf{Server side:} Enforce local DoS defense and support both Brotli and Gzip compression. 
After deploying a CDN, customer websites are under the well-advertised DDoS protections of the CDN. 
However, our CDN-Convert attacks can nullify this kind of protection. When suffering a CDN-Convert attack, the origin server can deploy a local DoS defense (e.g., filtering requests, limiting bandwidth, etc.) for temporary mitigation. 
But this does not necessarily work. 
From the perspective of the origin server, attack requests are no different from benign requests and come from widely distributed CDN nodes. 
It is difficult for the origin server to defend against it effectively without affecting normal services. 
To address CDN-Convert attacks at their root, it is advisable for the origin server to support both Brotli and Gzip compression.

\textbf{CDN side:} Modify the specific implementation of compression requests. 
In the HTTP context, CDNs do not support the Brotli compression algorithm, but they decompress gzip'd responses, which leads to the first type of CCUF attack. 
Therefore, OCDN can employ a strategy of not converting unsupported compression formats to defend against the first type of CCUF attack. 
UCDN's deletion strategy can result in the second type of CCUF attack. Therefore, CDNs can adopt the laziness strategy to fix the vulnerability.

\subsection{Responsible Disclosure}\label{subsec6.3}
We contacted all tested CDN vendors to responsibly disclose our findings, unveiling experiment details and reproducing procedures. 
They thanked our reports but have no further comments to date.

\section{Related Work}\label{sec7}

\subsection{HTTP Compression Security}\label{subsec7.1}
As far as we are aware, there hasn't been any academic research about the security threats posed by compression requests in CDN environments. While the CVE platform lists approximately ten vulnerabilities related to compression or decompression, it is worth noting that all of these vulnerabilities are associated with incorrect implementations but are unrelated to CDNs. For instance, CVE-2016-9843~\cite{CVE-2016-9843} resulted from buffer overflows due to Zlib's mishandling of deflate-compressed data. CVE-2018-1002105~\cite{CVE-2018-1002105} introduced a DoS attack leveraging the incorrect parsing of gzip-compressed data by the Kubernetes API. CVE-2023-3782~\cite{CVE-2023-3782} presented a scenario where an attacker could perform a man-in-the-middle attack and inject a Brotli compression bomb into an HTTP response, leading to client denial of service. Our CDN-Convert attacks primarily capitalize on the asymmetric traffic consumption in CDN frontend and backend connections, differing significantly from these vulnerabilities.

\subsection{CDN Security}\label{subsec7.2}
CDN is an important component of the internet infrastructure, has been extensively studied for its security~\cite{cdnoverall}. 
According to reports~\cite{reports}, nearly one-fifth of the current internet traffic is transmitted through CDNs.Therefore, the security of CDNs has always been a concern for researchers. For example, Triukose et al.~\cite{decoupleAmp}proposed an attack that exhausts the origin server bandwidth by quickly disconnecting the front-end connection. However, this attack has been proven ineffective in most CDNs. Additionally, Li et al.~\cite{RangeAmp} introduced the RangeAmp attack, which nullifies the DDoS protection of vulnerable CDNs and can be abused to attack the origin server. 
They also mentioned that attackers can set a small TCP receive window to receive only a small amount of data. 
In fact, we evaluated these attacks in our experiments and found that most tested CDNs can effectively mitigate them. 
Furthermore, CDNs can also become victims. In the forwarding loop attack proposed by Chen et al.~\cite{cdn-loop}, attackers can link CDN nodes in a loop, causing malicious requests to be repeatedly processed, thereby reducing the availability of the CDN. In contrast, our CDN-Convert attack presents a novel approach that can amplify attacks against both the CDN itself and the origin server, bypassing the DDoS protection provided by the CDN.

\subsection{Amplification Attacks}\label{subsec7.3}
Amplification attacks. Amplification attacks constitute a well-established area of research within the realm of cybersecurity. Booth et al.~\cite{UdpAmp} revealed that UDP amplification attacks, recruiting UDP servers on the internet as reflectors, can achieve amplification factors up to 556 times. Sieklik et al.~\cite{TFTP} conducted a further analysis of amplification attacks based on DNSSEC, resulting in amplification factors of up to 44 times. Beyond the UDP protocol, the TCP protocol is also susceptible to exploitation. Anagnostopoulos et al.~\cite{DnsAmp} studied TFTP amplification attacks with amplification factors of 60 times. Kuhrer et al.~\cite{TCPAmp}  delved into TCP reflection attacks across well-known TCP services like HTTP, MySQL, and POP3. Furthermore, Kuhrer et al.~\cite{NTP} revealed that NTP services could lead to amplification factors as high as 4670 times. Unfortunately, even the HTTP protocol has been abused. Li et al.~\cite{RangeAmp} exposed that amplification attacks leveraging the HTTP Range Request mechanism can achieve amplification factors as high as 43000 times. Guo et al.~\cite{CDNJudo} revealed that amplification attacks leveraging HTTP protocol conversion can achieve amplification factors as high as 166 times. Compared with these previous amplification attack studies, our attack can penetrate CDN DDoS protection and cause enormous amplification damage to the CDN-hidden website server and CDN nodes.

In conclusion, our research reveals that leveraging CDN compression mechanism can initiate a novel class of amplification attacks against websites hosted on CDNs and CDN nodes. These attacks undermine the DDoS protection offered by CDNs, presenting a significant threat to the internet security landscape.
\section{Conclusion}\label{sec8}

The CDN has undeniably become an indispensable part of the internet landscape, providing several benefits, such as acting as a shield against DoS attacks for websites hosted on CDN platforms. 
However, it is crucial to acknowledge that CDN's compression mechanisms can be used to exploit the undermining of the very DoS protection it is supposed to provide.

We have presented the principles of the CDN-Convet vulnerability, along with a comprehensive study of its practicality in the wild. 
We find that the 11 popular CDNs tested are all vulnerable. 
We believe that the CDN-Convet attacks will pose severe threats to the serviceability of CDNs and the availability of websites. We hope that our study will provide insight into this vulnerability and help the potentially relevant victims to fully understand them. 
We suggest that the CDNs and websites adopt one or more of the mitigation solutions discussed in our paper. 
As the CDN industry continues to expand and offer more features, we are committed to exploring these features in our future research.
\section{ACKNOWLEDGEMENTS}\label{sec9}
We thank the anonymous reviewers for their helpful feedback. 
This work is supported by the National Key Research and Development Program of China (Grant No. 2021YFB0301100)
\clearpage

\bibliographystyle{IEEEtran}
\bibliography{reference}

\end{document}